\documentclass[11pt]{article}
\usepackage{subfigure}
\usepackage{amsfonts}
\usepackage{times}
\usepackage{color}
\usepackage{graphicx,epsfig}
\usepackage{mdwlist}
\usepackage{wrapfig}
\usepackage{amssymb}
\usepackage{cite}      
\usepackage{amsmath}   
\usepackage{multirow}
\makeatletter

\newcommand{\RR}{\mathbb{R}}

\newcommand{\argmax}{\arg\max}

\newcommand{\commentout}[1]{}
\newcommand{\vect}[1]{\vec{#1}}

\newcommand{\singlespacing}{\let\CS=\@currsize\renewcommand{\baselinestretch}{1.0}\tiny\CS}
\newcommand{\oneandahalfspacing}{\let\CS=\@currsize\renewcommand{\baselinestretch}{1.2}\tiny\CS}
\newcommand{\doublespacing}{\let\CS=\@currsize\renewcommand{\baselinestretch}{1.4}\tiny\CS}
\makeatother

\oneandahalfspacing

\newcommand{\Xomit}[1]{}

\title{On Clustering on Graphs with Multiple Edge Types \thanks{ \footnotesize This work was funded by the applied mathematics program at the United States Department of Energy and performed at Sandia National Laboratories, a multiprogram laboratory operated by Sandia Corporation, a wholly owned subsidiary of Lockheed Martin Corporation, for the United States Department of Energy's National Nuclear Security Administration under contract DE-AC04-94AL85000. }}
\author{
\begin{tabular}{cc} Matthew Rocklin & Ali Pinar \\
Dept. Computer Science, U. Chicago &  Sandia National Laboratories \\
mrocklin@cs.uchicago.edu & apinar@sandia.gov  
\end{tabular} }

\date{}

\begin{document}
\maketitle
\begin{abstract} 
We study clustering on graphs with multiple edge types. Our main motivation is that similarities between objects can be measured in many different metrics. For instance similarity  between  
two papers can be based on  common authors,  where they are published, keyword similarity,  citations, etc.  As such, graphs with multiple edges  is a more accurate model to describe similarities between objects. Each edge/metric provides only partial information about the data; recovering full information requires aggregation of all the similarity metrics.  Clustering  becomes much more challenging in this context, since in addition to the difficulties of the traditional clustering  problem, we have to deal with a space of clusterings. We generalize the concept of clustering in single-edge graphs to multi-edged graphs and investigate  problems such as: Can we find a clustering that remains good, even if we change the relative weights of metrics? How can we describe the space of clusterings efficiently? Can we find  unexpected clusterings (a good clustering that is distant from all given clusterings)? If given the ground-truth clustering, can we recover  how the   weights for edge types  were aggregated? 

%
\end{abstract} 


%
\vspace{-5ex}
\section{Introduction}
\vspace{-2ex}
 A community or a cluster in a graph is a subset of vertices that are tightly coupled among themselves and loosely coupled with the rest of the graph.    Finding these communities is one of the fundamental problems of graphs analysis and  has  been the subject of numerous research efforts.   
 Most of these efforts begin with the premise that a simple graph is already constructed. A relation between two objects is represented by an edge between two nodes which may be weighted by the strength of the connection or left as a binary variable.  This paper studies the community detection problem on graphs with multiple edge types or multiple similarity metrics, as opposed to traditional graphs with a single edge type.   

In many real-world problems,  similarities between objects can be defined by  many different relationships. For instance,  similarity between two  scientific articles can be defined based on authors, citations to, citations from, keywords, titles, where they are published,  text similarity and many more.  Relationships between people can be based on the nature of the relationship (e.g., business, family, friendships) a means of communication (e.g., emails, phone, in person),
etc. Electronic files can be  grouped  by their type (Latex, C,  html),  names, the time they are created,  or the pattern they are accessed. In these examples, there are multiple graphs that define relationships  between the subjects.  
We may choose to reduce this multivariate information to construct a single composite graph. This is convenient as it enables application of many strong results from the literature. However, information being lost  during this aggregation may be crucial, and we believe working on graphs with multiple edge types is a more precise representation of the problem, and thus will lead to more accurate  analyses.  
Despite its importance, the literature on clustering graphs with multiple edge types is very rare. Mucha et al.\cite{Mucha2010} looked at community detection when multiple edge types are sampled in time and strongly correlated. Dunlavy et al. \cite{Dunlavy2006} described this problem as a three dimensional Tensor and used a PARAFAC decomposition (SVD generalization) to identify dominant factors.

The community detection problem on graphs with multiple edge types bears many interesting problems.  If a ground-truth clustering is known, can we recover an aggregation scheme that best resonates with the ground-truth data? How can we efficiently represent a space of clusterings spanned by the multiple edge types. Are the clusterings within this space clustered themselves? How do we find significantly different clusterings for the same data? 
These problems add another level of complexity to the already difficult problem of community detection in graphs.  As in the single edge type case,  the challenges lie not only in  algorithms, but also  in formulations of these problems.  In this paper, we investigate these problems and propose some solutions.  
Our techniques rely on using optimization (specifically methods that rely only on function values), methods for classical community detection (i.e., community detection with single edge types), and metrics for quantifying the distance between two clusterings.  We present results of three case studies on : 1)  file system data, where files are grouped to projects;  2) papers submitted to Arxiv, and 3)  countries  where  the data is based on CIA World Factbook~\cite{cia-nations}.

The rest of the paper is organized as follows.  In the next section, we will review some of the background information for this paper. In particular, we present the {\it variation of information} metric~\cite{Meila2003} that we use to quantify the distance  between two clusterings.   In Section~\ref{sec:groundtruth}, we study the problem of  computing an aggregate similarity measure for a given ground-truth clustering
Given a graph with multiple  edge types and the ground-truth clustering, can we  find  
an aggregation scheme to reduce the information from multiple edge types into a single metric, such that  this metric  best resonates with the ground-truth data?
We apply our methods to synthetic, and the file system data. 
Section~\ref{sec:latent}  addresses the latent clustering structure on graphs with multiple edge types, and introduces the concept of meta-clustering.  We also discuss  how we can represent the meta-clustering structure efficiently, and provide results on Arxiv data. In Section~\ref{sec:nations} we discuss how we can seek unexpected clusters and how we can improve the significance of clusters by  for multiple edge types.  We conclude with Section~\ref{sec:conc}.  

\section{Background}

A weighted graph is represented as a tuple $G = (V, E)$, $V$ a set of vertices and $E$ a set of edges. Each edge $e_i$ is a tuple $e_i \!=\! \{v_a, v_b, w_i \ |\  v_a,v_b \in V, w_i \in \RR\}$ representing a connection between vertices $v_a$ and $v_b$ with weight $w_i$. 
In this work we replace $w_i \!\in\! \RR$ with $\vect{w_i} \!\in\! \RR^k$ with $k$ being the number of edge types. We will refer to such graphs  as graphs with multiple edge types or multiweighted graphs.  
We will construct functions that map multiweighted edges $\vect{w_i}\in\RR^k$ to \textit{composite edge types} $f(\vect{w_i}) \!= \!\omega_i \!\in\! \RR$. For much of this paper $f$ will be linear $\omega_i \! =\! \sum \alpha_i w_i$.  

\subsection{Clustering} 

Intuitively, the goal of clustering is to break down the graph into smaller groups such that   vertices in each group  are tightly coupled among themselves and loosely coupled with the remainder of the graph.   Both the translation of this intuition into a well-defined mathematical formula and design of associated algorithms pose significant challenges.  
Despite the  high quality and the high volume of the literature, the area continues to draw a lot of interest due to the growing importance of the problem and the challenges posed by the size and mathematical variety of the subject graphs.   

Our goal is to extend the concept of clustering to graphs with multiple edge types without getting into the details of clustering algorithms  and formulations, since such a detailed study will be well beyond the scope of this paper. 
In this paper, we used {\it Graclus}, software developed by Dhillon et al\cite{Dhillon2007}, which uses  the top-down approach that recursively  splits the graph into smaller pieces and {\it FastCommunity} developed by Clauset et al\cite{Clauset2004} which uses an agglomerative approach which optimizes the modularity metric. For further information on clustering see Lancichinetti et al.\cite{Lancichinetti2009b}.

\subsection{Variation of information of clusterings}
\label{sec:background:VI}
At the core of most of our discussions will be  similarity between two clusterings, which calls for a method  to quantify the distance  between two clusterings. 
Several metrics and methods have been proposed for comparing clusterings, such as  {\it variation of information}~\cite{Meila2003}, {\it scaled coverage measure}~\cite{stichting00}, {\it classification error}~\cite{lange04,luo05,Meila2003}, and {\it Mirkin's metric}~\cite{mirkin96}.  Out of these, we have used the variation of information metric in our experiments.

Let $C_0=\langle C_0^1,C_0^2, \ldots ,C_0^K\rangle$ and $C_1=\langle C_1^1,C_1^2, \ldots ,C_1^K\rangle$ be two clusterings of the same node set.  
Let $n$ be the total number of nodes, and  $P(C, k)=\frac{|C^k|}{n}$ be the probability that a node is in cluster $C^k$ in a clustering $C$.  Similarly the probability that a node is in cluster $C^k$ in clustering $C_i$ and in cluster $C^l$ in clustering $C_j$ is $P(C_i,C_j,k,l)= \frac {|C_i^k\cap C_j^l|}{n}$. 
The {\it entropy of information} or expectation value of learned information in  $C_i$  is defined 
\[ H(C_i)=-\sum_{k=1}^K P(C_i,k)\log{P(C_i,k)}  \] 
the mutual information shared by $C_i$ and $C_j$ is 
\[ I(C_i,C_j)=\sum_{k=1}^K\sum_{l=1}^{K\prime}P(C_i,C_j,k,l)\log {P(C_i,C_j,k,l)},\]

Given these two quantities Meila defines the variation of information metric by
\begin{equation} 
\label{eq:vi1}
d_{VI}(C_i,C_j)=H(C_i)+H(C_j)-2I(C_i,C_j).  
\end{equation}
Meila~\cite{Meila2003} explains the intuition behind this metric a follows.  $H(C_i)$ denotes the average uncertainty of the position of a node in  clustering $C_i$.  If, however, we are given $C_j$, $I(C_i,C_j)$ denotes average reduction in uncertainty of where a node is located in $C_i$. If we rewrite Equation (\ref{eq:vi1}) as 
 \[ 
d_{VI}(C_i,C_j)=\left(H(C_i)-  I(C_i,C_j)\right)\;\; +\;\; \left(H(C_j)-  I(C_i,C_j)\right),
\]
the first term measures the information lost if $C_j$ is the true clustering and we know instead $C_i$, and the second term is the opposite. 
Note that $d_{VI}(C_i,C_j)$ will be  zero, when the two clusterings   are  the same, and it will be maximum when the two clusterings are independent.
 
The variation of information metric can be computed in $O(n)$ time.

\section{Recovering a graph given a ground truth clustering}
\label{sec:groundtruth}


Suppose we are given a ground-truth clustering for a graph with multiple edge types/similarity metrics.
Can we recover an aggregation scheme that best resonates with the ground-truth data?  This aggregation scheme that reduces multiple similarity measurements into a single similarity measurement can be  a crucial enabler that reduces the problem of  finding communities with multiple similarity metrics,  to a well-known, fundamental problem in data analysis. Additionally if we can obtain this aggregation scheme from data sets for which the ground-truth is available, we may then apply the
same aggregation to other data instances in the same domain. 

Formally,  we work on the following problem. Given a graph $G=(V,E)$ with multiple  similarity measurements for each edge $\langle w_i^1,w_i^2,\ldots ,w_i^K\rangle \in \RR^K$, and a ground-truth clustering for this graph $C^*$. 
Our goal is to find a weighting vector $\alpha\in \RR^K$, such that the $C^*$ is an optimal clustering for the graph $G$, whose edges are weighted as $w_i=\sum_{j=1}^K\alpha_jw^j_i$.   Note that this is only a semi-formal definition, as we have not formally defined what we mean by an {\it optimal clustering}.   In addition  to the well-known difficulty of  defining what a good clustering means, matching to the ground-truth data has specific challenges, which  we discuss in the subsequent section. 
 
Below, we describe two approaches. The first  approach  is based on inverse problems, and we try to find weighting parameters for which  the clustering on the graph yields the ground-truth clustering.  The second approach computes weighting parameters that maximizes the quality of the ground-truth clustering. 

\subsection{ Solving an inverse problem}

Inverse problems arise in many scientific computing applications where  the goal is  to infer unobservable parameters from finite observations.  Solutions typically  involve iterations of predictions   and  solution of  the {\it  forward} problems to compute the accuracy  of the prediction.  
Our problem can be considered as an inverse problem, since we are trying to  compute an aggregation function, from  a given clustering.  The forward problem in this case is  the clustering operation.  We can start with a random guess for the edge weights, cluster the new graph, and use the distance between two clusterings as a measure for the quality of the guess. We can further put this process within an optimization loop to find the parameters that yield  the closest clustering to the ground-truth.  

The disadvantage of this method is that it relies on the accuracy of the forward solution, i.e., the clustering algorithm.  If we are given the true solution to the problem, can we construct the same clustering?  This will not be easy for two reasons. 
First,  there is no perfect clustering algorithm, and secondly, even if we were able to solve the clustering problem optimally, we would not  have the exact objective function for clustering. 
Also, solving many clustering problems will be time-consuming especially for large graphs.  

\subsection{Maximizing the quality of ground-truth clustering}
An  alternative approach is to find an aggregation function that maximizes the quality of the ground-truth clustering.
For this purpose, we have to take into account not only  the overall quality of the clustering, but also the  placement of individual vertices, as the individual vertices represent local optimality.  For instance, if the quality of the clustering will improve by moving a vertex to another cluster than its  ground-truth, then the current solution  cannot be ideal.  While it is fair to assume some vertices might have been misclassified in the ground-truth data, there should be a penalty for such vertices. Thus we have two objectives while computing $\alpha$: {\it (i)} justifying the location of each vertex  
{\it (ii)}  maximizing the overall quality of the clustering. 
%

\subsubsection{Justifying locations of individual vertices}

For each vertex $v\in V$ we define the \textit{pull} to each cluster $C^k$  in $C=\langle C^1,C^2, \ldots  C^K  \rangle $ to be the cumulative weights of edges between $v$ and its neighbors in $C^k$, 
\begin{equation}
P_\alpha(v,C_k) = \sum_{w_i=(u,v)\in E; u\in C^k} w_i(\alpha)
\end{equation}
We further define the \textit{holding power}, $H_\alpha(v)$ for each vertex, to be the pull of the cluster to which the vertex belongs in $C^*$ minus the next largest pull among the remaining clusters. If this number is positive then $v$ is held more strongly to the proper cluster than to any other. 
We can then maximize the number of vertices with positive holding power by maximizing $ |\{v : H_\alpha(v)>0\}|$.
What is important here is the concept  of pull and hold, as the specific definitions may be changed without altering the core idea.

\begin{wrapfigure}{l}{0.45\textwidth}
\vspace{-2.5ex}
\hspace*{-5ex}
\includegraphics[width=.5\textwidth]{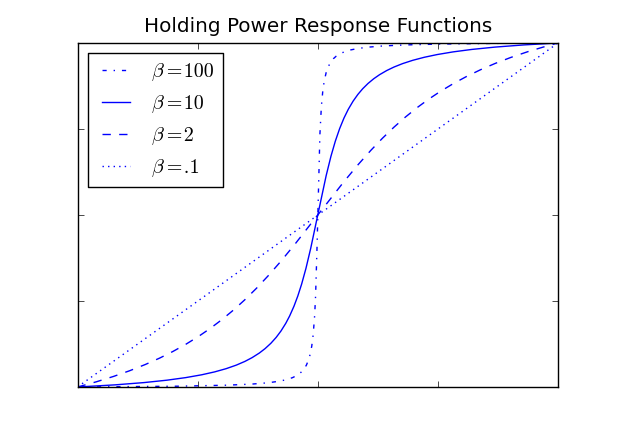}
\vspace{-4.5ex}
\caption{\label{fig:responsefunctions} \small Arctangent provides a smooth blend between step and linear functions. }
\vspace{-3ex}
\end{wrapfigure}
While this method is local and easy to compute, its discrete nature limits the tools that can be used to solve the associated optimization problem. Because gradient information is not available it hinders our ability to navigate in the search space.  In our experiments, 
we smoothed the step-like nature of the function $H(v)>0$ by replacing it with $\arctan(H_\alpha(v))$. This functional form
 still encodes that we want holding power to be positive for each node but it allows the optimization routine to benefit from small improvements. It emphasises nodes which are close to the $H(v)=0$ crossing point (large gradients) over nodes which are well entrenched (low gradients near extremes).

This objective function sacrifices holding scores for nodes which are safely entrenched in their cluster (high holding power) or are lost causes (very low holding power) for those which are near the cross-over point. The extent to which it does this can be tuned by the steepness parameter of the arctangent.
  For very steep parameters this function resembles classification (step function) while for very shallow parameters it resembles a simple linear sum as seen in
Fig.~\ref{fig:responsefunctions}. 
 We can solve the following optimization problem to maximize the number of vertices, whose positions in the ground-truth clustering are justified by the weighting vector $\alpha$.  
\begin{equation}
\argmax_{\alpha\in\RR^K} \sum_{v\in V} \arctan(H_\alpha(v))
\end{equation}


\subsubsection {Overall clustering quality}
In addition to individual vertices  being justified, overall quality of the clustering should be  maximized, as well.  Any quality metric can potentially be used for this purpose however we find that some strictly linear functions have a trivial solution. Consider an objective function that measures the quality of a clustering as the sum of the inter-cluster edges. 
To minimize the cumulative weights of cut edges, or equivalently to maximize the cumulative weights of internal edges we solve
\begin{eqnarray*} 
 \min_\alpha &   \sum_{e_{j}\in Cut}\sum_{i=1}^K \alpha_i w^i_j \;\;\; |\alpha|=1\\
\end{eqnarray*}
where $Cut$ denotes the set of edges whose end points are in different clusters. 
Let $S^k$ denote the sum of the cut edges with respect to the $k$-th metric.
 That is $S^k=\sum_{e_{j}\in Cut} w_j^k$.  Then the objective function can be rewritten as $\min_\alpha \sum_1^K \alpha_kS^k$. Because this is linear it has a trivial solution that assigns 1 to the  weight of the maximum $S^k$, which means  only one similarity metric is taken into  account. 
While additional constraints may exclude this specific solution, a linear formulation of the quality will always yield only a trivial solution within the new feasible region.  

In our experiments we used the {\it modularity}  metric~\cite{modularity}.  The modularity metric  uses a random graph  generated with respect to the degree distribution as  the null hypothesis, setting the modularity score of a random clustering to 0.  Formally,  the modularity score for an unweighted graph  is 
\begin{equation}
\frac{1}{2m}\sum_{ij} \left[e_{ij} - \frac{d_id_j}{4m}\right]\delta_{ij},
\end{equation}
where $e_{ij}$ is a  binary variable that is 1, if and only if  vertices $v_i$ and $v_j$ are connected;  $d_i$ denotes the degree of vertex $i$; $m$ is the number of edges; and $\delta_{i,j}$  is a  binary variable that is 1, if and only  if  vertices $v_i$ and $v_j$ are on the same cluster.   In this formulation,  $\frac{d_id_j}{4m}$ corresponds to the number of edges between vertices $v_i$ and $v_j$ in a random graph with the given degree distribution, and its subtraction corresponds to the  the null hypothesis.

This formulation can be generalized for weighted graphs by redefining $e_{ij}$ as  the weight of this edge (0 if no such edge exists), $d_i$ as  the cumulative weight of edges incident to $v_i$; and $m$ as the cumulative weight of all edges in the graph~\cite{wmodularity}.

\subsubsection{Solving the optimization problems }
\label{sec:opt}

We have presented several nonlinear optimization problems for which the derivative information is not available. 
 To solve these problems we used HOPSPACK (Hybrid Optimization Parallel Search PACKage)~\cite{hopspack}, which is developed at Sandia National Laboratories to solve linear and nonlinear optimization problems  when the derivatives are not available.  

\subsection{Experimental results}
%
\subsubsection {Recovering edge weights}
 The goal of this  set of experiments is  to see whether we can find aggregation functions that justify a given clustering. We have performed our experiments on 3 data sets.

\noindent{\bf  Synthetic data:}
Our generation method is based on Lancichinetti et al.'s work \cite{Lancichinetti2008} that proposes a method to generate graphs as benchmarks for clustering algorithms.
We generated graphs of sizes 500, 100, 2000, and 4000 nodes, 30 edges per node on average, mixing parameters $\mu_t=.7, \mu_w=.75$, and known communities. We then perturbed edge weights, $w_i$, with additive and multiplicative noise so that $w_i\leftarrow \nu(w_i+\sigma) : \sigma\in (-2w_{a}, 2w_{a}), \nu \in(0,1)$ uniformly, independently, and identically distributed, where $w_{a}$ is the average edge weight.

\begin{figure}[bth]
\centering
\includegraphics[width=1.0\textwidth]{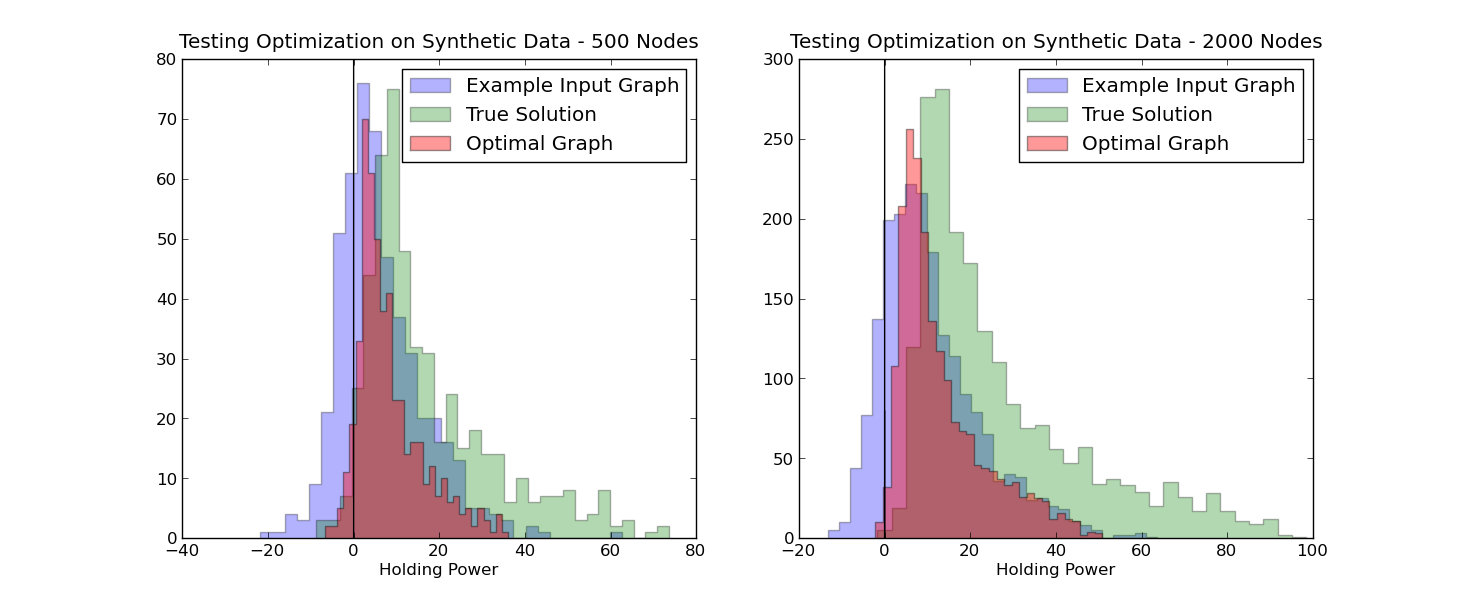}
\caption{\label{fig:lfrhist} \small Three histograms of holding powers for Blue: an example perturbed (poor) edge type, Green: the original data (very good), Red: the optimal blend of ten of the perturbed edge types. }
\end{figure}

After the noise, none of the original metrics  preserved the original clustering structure. We display this in Fig.~\ref{fig:lfrhist}, which  presents histograms for the holding power for vertices. The green bars correspond to vertices of the original graph, they all have positive holding power.  The blue bars correspond to holding powers after  noise is added. We only present  one edge type for  clarity of presentation. As can be seen a significant portion
(30\%) of the vertices have negative holding power, which means  they would rather be on another cluster.  The red bars show the holding powers after we compute an optimal linear aggregation. As seen in the figure, almost all vertices move to the positive side, justifying the ground-truth clustering.  A few vertices with negative holding power are expected, even after an optimal solution due to the noise. These  results show that a composite similarity  that resonates with
a given clustering can be computed  out of many metrics, none of which give a good solution by itself.  

In Table~\ref{tab:lfrtest},  we present these results on graphs with different number of vertices.  While the percentages change for different number of vertices, our main conclusion that a good clustering can be achieved  via a better  aggregation function remains valid. 

\begin{table}[tbh]

\caption{ \label{tab:lfrtest} \small Fraction of nodes with positive holding power for  ground-truth, perturbed, and optimized graphs}
\vspace{-2ex}
\begin{center}
  \begin{tabular}{| c | c | c | c | c | }
    \hline
    Number of nodes &Number of clusters&  Ground-truth & Optimized  & Perturbed (average) \\ \hline
    500 &  14  & .965 & .922 & .703 \\
    1000 & 27 & .999 & .977 & .799 \\
    2000 & 58 &  .999 & .996 & .846 \\
    4000 & 118& 1.00 & .997 & .860 \\
    \hline
  \end{tabular}
\end{center}
\end{table}

\noindent
{\bf  File system data:} 
An owner of a  workstation classified  300 of his files as belonging to one of his three ongoing projects, which we took as the ground-truth clustering.  We used  filename similarity, time-of-modification/time-of-creation similarity, ancestry (distance in the directory-tree), and parenthood (edges between a directory node with file nodes in this directory) as the similarity metrics among these files. 

\begin{figure}[thb]
\centering
\includegraphics[width=0.6\textwidth]{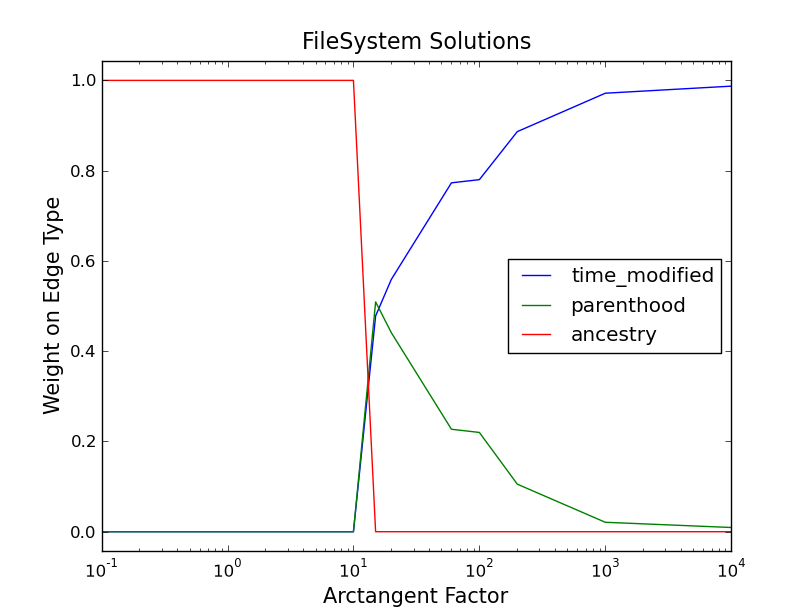}
\caption{\small Optimal solutions  for the file system data for different arctangent parameters} 
\label{fig:filesystem}
\end{figure}
Our  results showed that only three metrics (time-of-modification, ancestry, and parenthood) affected the clustering.  However, the  solutions were sensitive to the choice of the arctangent parameter. In Fig.~\ref{fig:filesystem}, each column corresponds to an optimal solution for the corresponding  arctangent parameter.  Recall that  higher values of  the arctangent parameter corresponds to sharper step functions. Hence, the right of the figure corresponds to  maximizing the total number of  vertices with positive holding power, while the left side corresponds to  maximizing the sum of holding powers. The difference between the two is that the solutions on the right side may have a lot of nodes with barely positive values, while those on the left  may have nodes further away from zero at the cost of  more nodes with negative holding power. 
This is expected in general, but drastic change in the optimal solutions  as we go from one extreme to another was surprising to us, and should be taken into account in further studies. 

\noindent{\bf Arxiv data:}
We  took  30,000 high-energy physics articles published on arXiv.org and considered abstract text similarity, title similarity, citation links, and shared authors as edge types for these articles. We used  the top-level domain  of the submitter's e-mail (.edu, .uk, .jp, etc...) as a proxy for the region where the work was done. We used these regions as the ground-truth clustering.

The best parameters  that  explained the  ground-truth clustering were  0.0 for  abstracts, 1.0 for authors, 0.059  for citations, and 0.0016 for titles.  This means the  shared authors edge type is almost entirely favored, with cross-citations coming a distant second. This is intuitive because a graph of articles linked by common authors will be linked both by topic (we work with people in our field) but also by geography (we often work with people in a nearby institutions) whereas edge types like abstract text similarity tend to encode only the topic of a paper, which is less geographically correlated.  Different groups can work on the same topic, and  it was good to see that citations  factored in, and such a clear dominance of the authors information was noteworthy.  As a future work, we plan to investigate  nonlinear aggregation functions on this graph. 

\subsubsection{Clustering quality vs.  holding vertices } 
\begin{figure}[t]
\centering
\includegraphics[width=.55\textwidth]{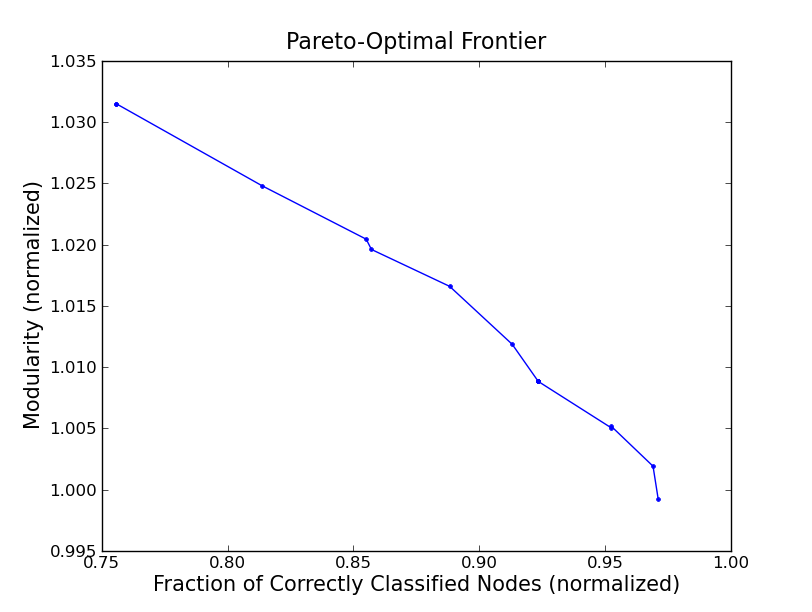}
\caption{\small Pareto frontier for two objectives: normalized modularity  and percentage of nodes  with positive holding power}
\label{fig:pareto}
\end{figure}

We have stated two goals  while computing an aggregation function: justifying the position of each vertex and the overall quality of clustering.  
In Fig.~\ref{fig:pareto}, we present the Pareto frontier for the two objectives.  The vertical axis  represent the quality of the clustering with respect to the modularity metric~\cite{modularity}, while the horizontal axis represents the percentage of nodes with positive holding power.  The  modularity numbers are normalized with respect to the modularity of the ground-truth clustering, and  normalized  numbers can be above 1, since  the ground-truth clustering does not specifically aim at maximizing modularity. 

As expected, Fig.~\ref{fig:pareto}  shows a trade-off between two objectives.  However, the scale difference between the two axis should be noted.  The full range in modularity change is limited to only 3\% for modularity, while the range is more than 20\% for  percentage of vertices with positive holding power. More importantly,  by only looking at the holding powers we can be preserve the original modularity quality.   The reason for this is that we have relatively small clusters, and almost all vertices
 have a connection with a cluster besides their own. If we had clusters where  many vertices had all their  connection within their clusters (e.g., much larger clusters), then this would not have been the case, and having a separate quality of clustering metric would have made sense.  However, we know that most complex networks have small communities no matter how big the graphs are~\cite{leskovec:conductance}.  Therefore, we expect that looking only at the  holding powers of vertices will be sufficient to recover aggregation functions.  

 \subsubsection{Inverse problems  vs.  maximizing clustering quality} 
\begin{figure}[t]
\centering
\includegraphics[width=.6\textwidth]{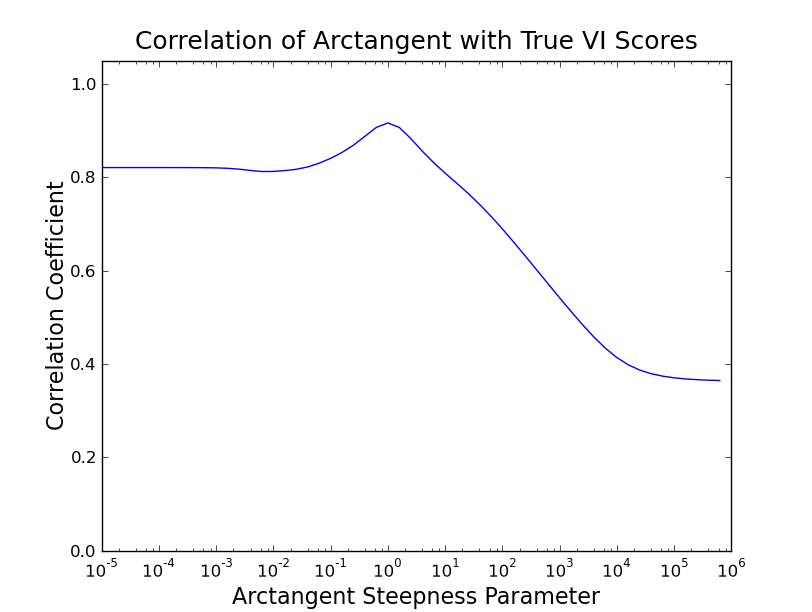}
\caption{\small The correlation of the arctan-smoothed objective function with variation of information distance using clusterings generated by Graclus  as we vary the steepness parameter} 
\label{fig:correlation}
\end{figure}

We used the file system data set to investigate the relationship between the two proposed approaches, and we present our results in Fig.~\ref{fig:correlation}.  For this figure, we computed the objective function for the ground-truth clustering for various aggregation weights and used the same weights to compute clusterings  with Graclus. From these clusterings we computed the variation of information (VI) distance to the ground-truth.
Fig.~\ref{fig:correlation} presents the correlation between the  measures:  VI distance for Graclus clusterings for the first approach, and the objective function values for the  second approach.  
This tries to  answer whether solutions with higher  objective function values yield clusterings closer to the ground-truth using Graclus.   In this figure,  a horizontal line fixed at 1 would have been ideal, indicating perfect correlation.   Our results nevertheless, show a very strong correlation between the two.   These results are not sufficient  to be conclusive as we need more experiments and other clustering tools.  However, this experiment produced promising results and  shows how such a study may be performed.   

\subsubsection{ Runtime scalability}
\begin{figure}
\centering
\includegraphics[width=.5\textwidth]{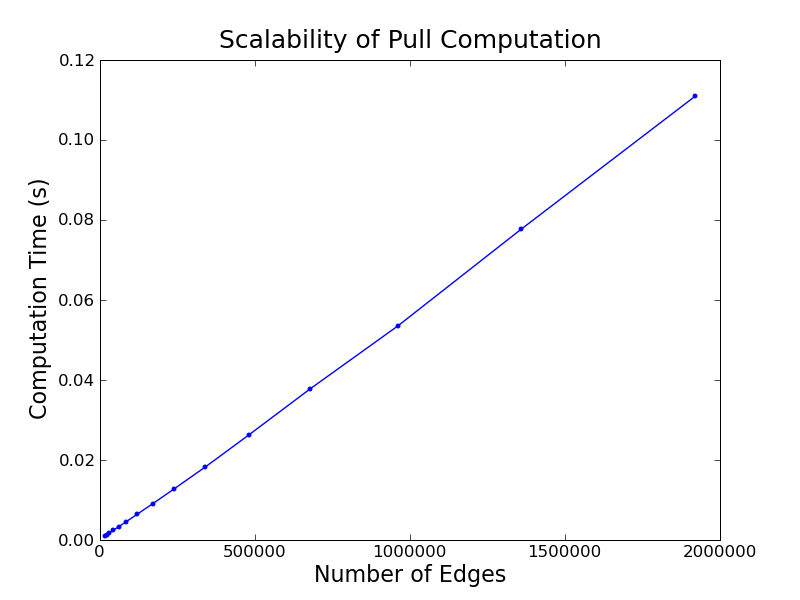}
\caption{\small Scalability of the proposed method }
\label{fig:runtime}
\end{figure}

In our final set of experiments we show the scalability of the proposed method. First, we want to note that  the number of unknowns  for the optimization problem is only a function of the aggregation function and is independent of the  graph size.  The required number of operations for one function evaluation on the other hand  depends linearly on the size of the graph, as illustrated in Figure~\ref{fig:runtime}.  In this experiment, we used  synthetic graphs with 30 as the average degree, and the presented numbers correspond to averages on 10 different runs.   As expected, the runtimes scale linearly with the number of edges.   
     
The runtime of the optimization algorithm depends  on the number of function evaluations. Since the algorithm we used is nondeterministic, the number of function evaluations, hence runtimes vary  even for different runs on the same problem, and thus are less informative. We are not presenting these results in detail due to space constraints.  However, we want to reemphasize that the size of the optimization problem  does not grow with  the graph size, and we don't expect the number of functions evaluations to cause any scalability problems.   

 We also observed that the number of function evaluations increase linearly with the number of similarity metrics.  These results are also omitted due to space constraints. 

\section{ Finding latent clusters}  
\label{sec:latent}
Consider the situation where several edge types share redundant information yet as an ensemble combine to form some broader structure. For example scientific journal articles can be connected by text similarity, abstract similarity, keywords, shared authors, cross-citations, etc.... Many of these edge types reflect the \textit{topic} of the document while others are also influenced by the \textit{location} of the work. 
Text, abstract, and keyword similarity are likely to be redundant in conveying topic information (physics, math, biology) while shared authorship (two articles sharing a common author) is likely to convey both topic and location information because we tend to work with both those in our same field and with those in nearby institutions. We say that the topic and location attributes are \textit{latent} because they do not exist explicitly in the data.
We can represent much of the variation in the data by two relatively independent clusterings based on the topic of documents and their location. This compression of information from five edge types to two meaningful clusterings is  the goal of this paper.

\subsection{An illustrative problem}
\begin{figure}[ht]
\centering
  \subfigure[270 vertices arranged in nine clusters on the plane. Edges exist between vertices so that close points are well connected and distant points are poorly connected.]
  {
    \includegraphics[width=.4\textwidth]{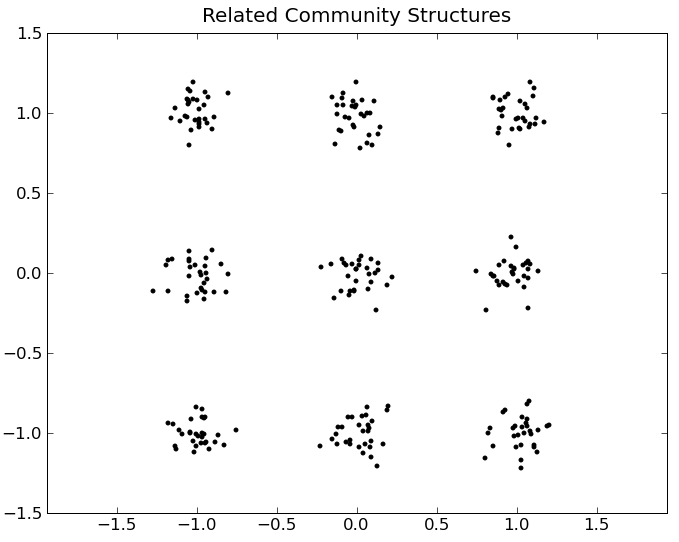} 
  }
  \hspace{.05\textwidth}
  \subfigure[Two 1D graphs arranged to suggest their relationship to the underlying 3x3 community structure. Both have clear community structures that are related but not entirely descriptive of the underlying 3x3 communities. ]
  {
    \includegraphics[width=.4\textwidth]{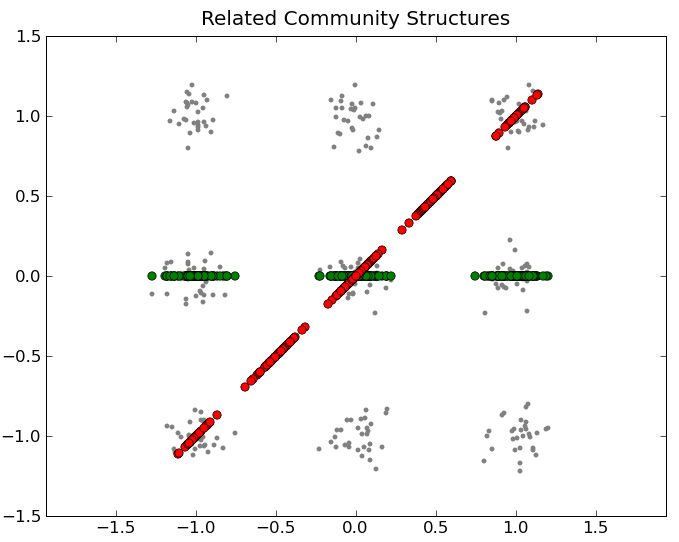}
  }
\caption{\small Illustrating clusters  (a) underlying structure  and (b) low-dimensiona/partial views } \label{fig:3x3}
\end{figure}

We construct a graph with multiple edge types  to demonstrate latent classes. For illustration, we assume our graph is perfectly embedded in  $\RR^2$ as seen in Fig. ~\ref{fig:3x3}a.  
In this example each point on the plane represents a vertex, and  two vertices  are connected by an edge if they are close  in distance.  The similarity/weight for each edge is inversely proportional to the Euclidean distance. We see visually that there are nine natural clusters. More interestingly we see that these clusters are arranged symmetrically along two axes. These clusters have more structure than the set $\{1,2,3,...,9\}$. Instead they have the structure $\{1,2,3\} \times \{1,2,3\}$. An example of such a structure would be the separation of academic papers along two factors, \{Physics, Mathematics, Biology\} and \{West Coast, Midwest, East Coast\}. The nine clusters (with examples like physics articles from the West or biology articles from the Midwest) have underlying structure. 

Our data sets do not directly provide this information. For instance with journal articles we can collect information about authors, where the articles are published, and their citations.  Each of these aspects provides only a partial view of the underlying structure.  Analogous to our geometric example above we could consider  features of the data as projections of the points to one dimensional subspaces. Distances/similarities between the points in a projection have only  partial information. This is depicted pictorially in Fig. ~\ref{fig:3x3}b. For instance, the green projection represents a metric that clearly distinguishes between columns but cannot differentiate between different communities on the same column. The red projection on the other hand provides a diagonal sweep,  capturing partial information about columns and partial information about rows.  
Neither of the two metrics can provide the full information for the underlying data. However when considered as an ensemble they do provide a complete picture.  Our goal is to be able to tease out the latent factors of data from a given set of partial views.

In this paper, we will use this $3\times 3$ example  for conceptual purposes and for illustrations.  
Our approach is  construct  many multi-weighted graphs by using combinations of the partial views of the data. We will cluster these graphs and analyze these clusters to recover the latent structure.

We expect that different regions of this space will have different clusterings. How drastic these differences are will depend on the particular multiweighted graph. How can we characterize this space of clusterings? 
Are there homogeneous regions, easily identifiable boundaries, groups of similar clusterings, etc...? We investigate the existence of a meta-clustering structure. That is we search for whether or not several clusterings in this space exhibit community structure themselves. In this section, we present our  methods  for these questions on the $3\times 3 $ data.   We will later provide results on a larger data set.  

\subsection{Sampling the clustering space}

To inspect the space of clusterings we sample in a Monte Carlo fashion. We take points $\vect{\alpha_i} \in \RR^k$ such that $|\vect{\alpha_i}|=1$, and compute the appropriate graph and clustering at each point. We may then compare these clusterings in aggregate. 

As our first experiment,  we take 16 random one-dimensional projections of the points laid out in the plane shown in Fig. ~\ref{fig:3x3} and consider the projected-point-wise distances in aggregate as a multiweighted graph. From this multiweighted graph we take $800$ samples of the linear space of clusterings. These 800 clusterings approximate the clustering structure of the multiweighted graph. 

\begin{figure}[ht]
\centering
  \subfigure[]
  {
    \label{fig:VIclusteringDistances}
    \includegraphics[width=.44\textwidth]{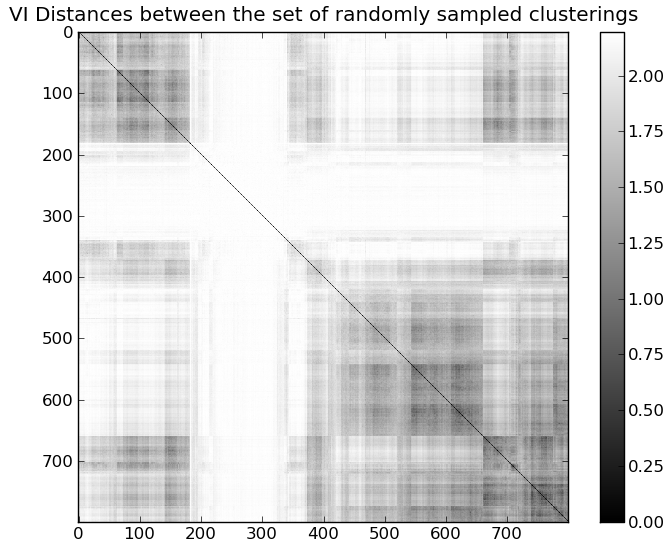} 
  }
  \hspace{.1\textwidth}
  \subfigure[]
  {
    \includegraphics[width=.28\textwidth]{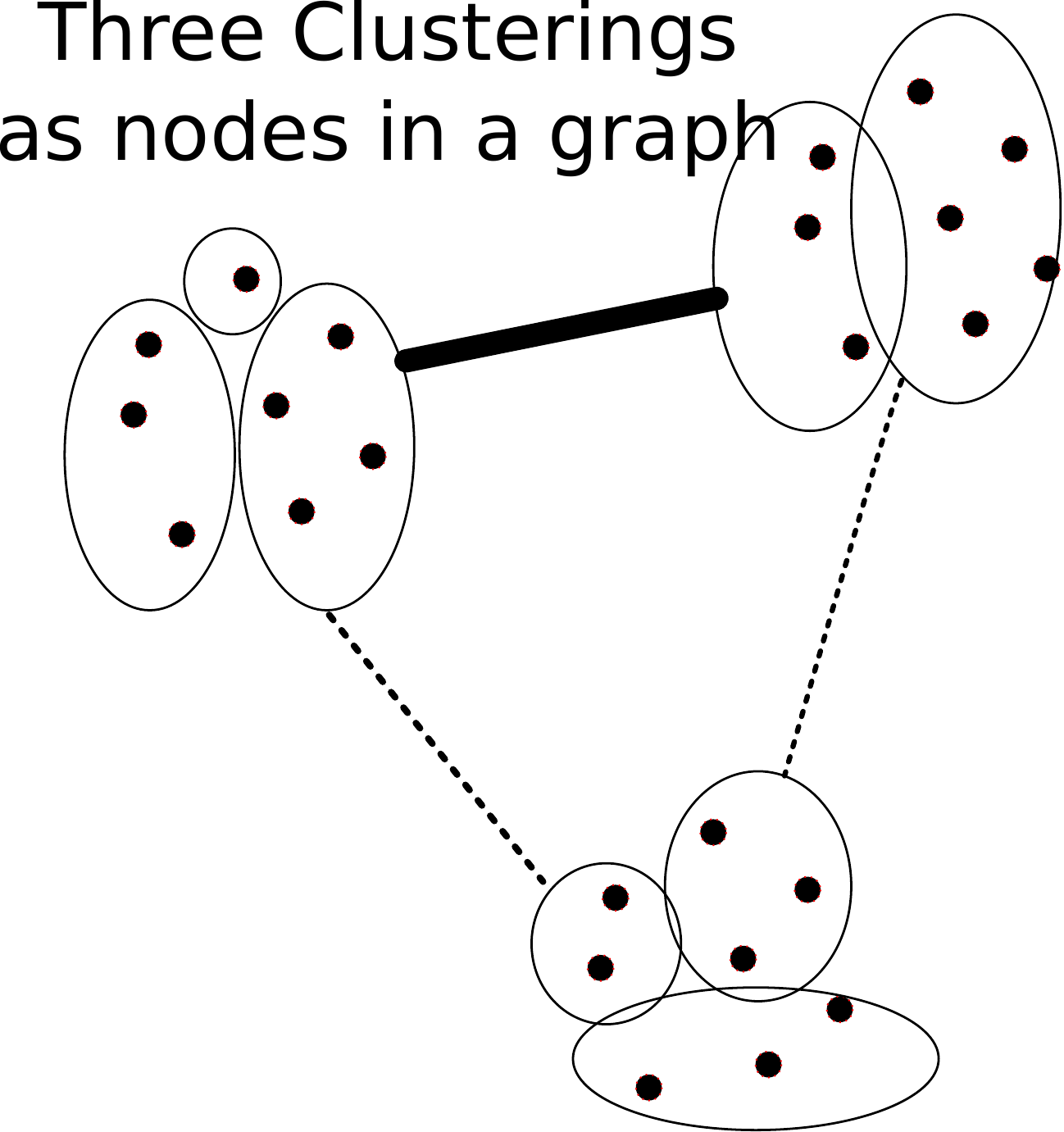}
    \label{fig:clusteringsGraph}
  }
\caption{\small The Meta-clustering  information (a) VI distances between 800 sampled clusterings. Vertices are ordered to show optimal clustering of this graph. Dark blocks on the diagonal represent clusters. The white band is a group of completely independent clusterings.  (b) Three Clusterings treated as nodes in a graph. Similar clusterings (top two) are connected with high-weighted edges. Distant clusterings are connected with low-weighted edges.} 
\end{figure}

The results of these experiments are presented in Figure~\ref{fig:VIclusteringDistances}. In this figure each row and column corresponds to a  clustering of the graph. Entries in the matrix represent  the variation of information distance between two clusterings. Therefore dark regions in this matrix are sets of clusterings that are highly similar. White bands show informational independence between regions. The rows/columns of this matrix have been ordered to have more similar clusterings closer to each other so as to highlight the clusters of clusterings
detected.

\vspace{-3ex}
\subsection{Meta-clusters: clusters of clusterings}
\label{sec:clustersOfClusterings}
While it is interesting to know that significantly different clusterings can be found, the lack of  stable clustering structure is not helpful for applications of clustering such as  for unsupervised learning. We need to reduce this set of clusterings further. We approach this problem by applying the idea of clustering onto this set of clusterings. We call this problem the {\it meta-clustering} problem. 

We represent the clusterings as nodes in a graph and connect them with edge-weights determined by the inverse of the variation of information metric \cite{Meila2003}. We inspect this graph to see if it contains clusters. 
That is, we \textit{cluster the graph of clusterings} to see if there exist some tightly coupled clusters of clusterings within the larger space. For instance in Fig. ~\ref{fig:clusteringsGraph} the top two clusterings differ only in the position of a single vertex and thus are highly similar. In contrast the bottom clustering is different from both and is weakly connected.

Figure~\ref{fig:VIclusteringDistances}  reveals the meta-clustering structure in our experiments.  The dark blocks  around the diagonal correspond to meta-clusters.  We can see two big blocks in the upper left and lower right corners.  Furthermore, there is a hierarchical clustering structure within these blocks, as we see smaller blocks within the larger blocks. 
In this experiment, we were able to observe meta-clusters. As usual,  results  depend on the particular problem instance.  While we do not claim that one can always find such  meta-clusters, we expect that  they will exist in  many multi-weighted graphs, and exploiting the meta-clustering structure can enable  efficiently handling this space, which is the topic of the next section.

\subsection{Efficient representation of the clusterings} 
In this section we study how to efficiently represent the meta-clustering structure. First we will study how to reduce a cluster of clusterings into a single averaged or representative clustering.  Then, we will study how to select and order a small number of meta-clusters to cover the clustering space efficiently.

\subsubsection{Averaging clusterings within a cluster}
\begin{figure}[ht]
\vspace{-8mm}
\centering
\includegraphics[width=.99\textwidth]{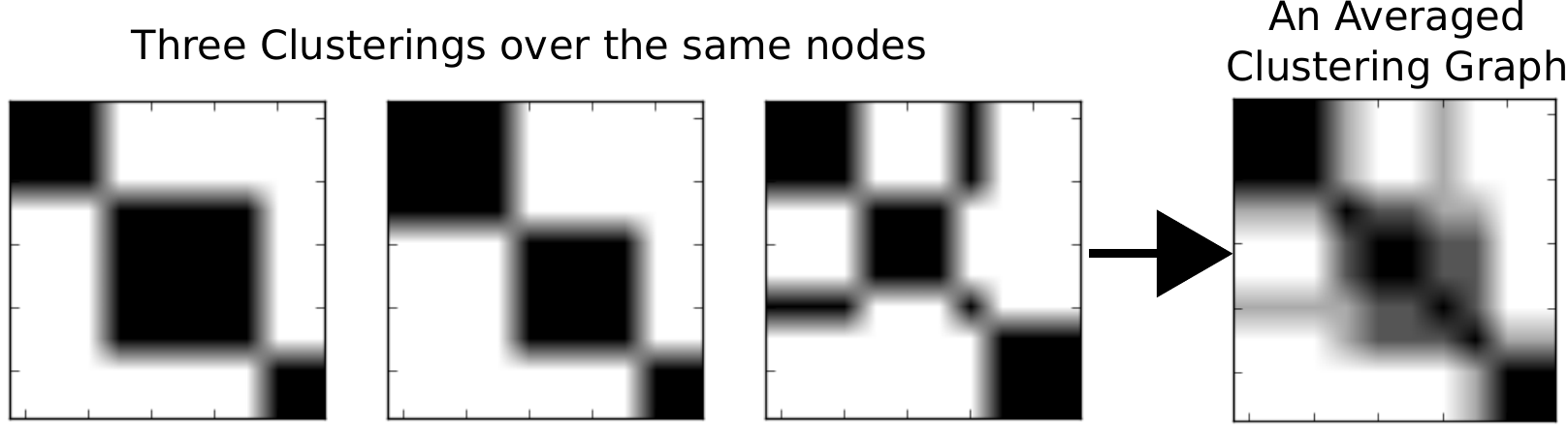}
\vspace{-0pt}
\caption{\small Showing the CSPA \cite{Strehl2003} averaging procedure for clusterings. Each clustering is displayed as a block diagonal graph (or permutation) with two nodes connected if and only if they are in the same cluster. Then an aggregate graph (right) is formed by the addition of these graphs. This graph on the right is then clustered using a traditional algorithm. This clustering is returned as the representative-clustering.}
\label{fig:CSPA}
\vspace{00pt}
\end{figure}

To increase the human accessibility of this information we reduce each cluster of clusterings into a single representative clustering.  We use the "Cluster-based Similarity Partitioning Algorithm" (CSPA) proposed by Strehl et. al \cite{Strehl2003}  to combine several clusterings into a single average. In this algorithm each pair of vertices is connected with an edge with weight equal to the number of clusters in which they co-occur. If $v_a$ and $v_b$ are in the
same cluster in $k$ of the clusterings then in this new graph they are connected with weight $k$. If they are never in the same cluster then they are not connected. We then cluster this graph and use the resultant clustering as the representative. In Fig. ~\ref{fig:CSPA} we depict the addition of three clusterings to form an average graph which can then be clustered.

\begin{figure} 
\centering
  \subfigure
  {
    \includegraphics[width=.23\textwidth]{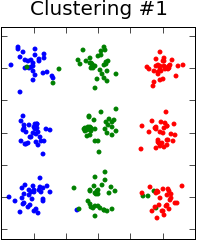} 
  }
  \subfigure
  {
    \includegraphics[width=.23\textwidth]{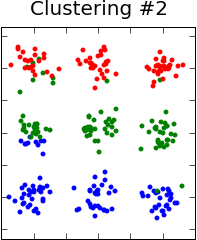}
  }
  \subfigure
  {
    \includegraphics[width=.23\textwidth]{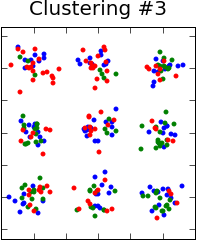} 
  }
  \subfigure
  {
    \includegraphics[width=.23\textwidth]{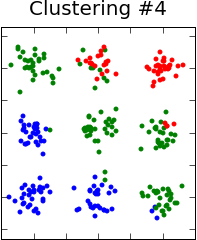}
  }
\caption{\small Representative-Clusterings of the four dominant clusters-of-clusterings from Fig. ~\ref{fig:VIclusteringDistances}. Clusterings are displayed as colorings of the original points in the 2-d plane. These are ordered to maximize cumulative set-wise information. Notice how the first two representative-clusterings recover the original nine clusterings exactly. } 
\label{fig:clusteringRepresentatives}
\end{figure}

We perform this process on the clusters of clusterings found in section \ref{sec:clustersOfClusterings} and presented in Fig. ~\ref{fig:VIclusteringDistances} to obtain the \textit{representative-clusterings} in Fig.~\ref{fig:clusteringRepresentatives}. We see that the product of the first two representative-clusterings identifies the original nine clusterings with little error. We see also that the two factors are identified perfectly by each of these clusterings
individually. 

\vspace{-3ex}
\subsubsection{Ordering by set-wise information content}
\label{sec:orderingInformationContent}
In Fig. \ref{fig:clusteringRepresentatives}, the original 3x3 community structure  can be reconstructed using only the first two representative-clusterings. Why are these two chosen first? Selecting the third and fourth representative-clusterings would not have had this pleasant result. How should we order the set of representative-clusterings?

We may judge a set of representative-clusterings by a number of factors: {\it (i)}  How many of our samples ascribe to the associated meta-clusters, what fraction of the space of clusterings do they cover? 
{\it (ii)} How much information do the clusterings cover as a set? 
{\it (iii)} How redundant are the clusterings? How much informational overlap is present? 
We would like to maximize information while minimizing redundancy. In Fig. \ref{fig:clusteringRepresentatives} we ordered the representative-clusterings to maximize setwise information. 
Minimizing redundancy came as a fortunate side-effect. Notice how each of the clusterings in order is independent from the preceding ones. Knowing that a vertex is red in the first image tells you nothing about the color of the vertex in the second. The second therefore brings only novel information and no redundancy. 

To compute the information content of a set of clusterings we extend the Variation of Information metric in a natural way. In Section \ref{sec:background:VI} we introduced the mutual information of two clusterings as follows:\\
\[ I(C_i,C_j)=\sum_{k=1}^K\sum_{l=1}^{K\prime}P(C_i,C_j,k,l)\log {P(C_i,C_j,k,l)},\]
where  $P()$ is the probability that a randomly selected node was in the specified clusters. This is equivalent to the self-information of the Cartesian product of the two clusterings. Its extension to a set of clusterings  $I(C_\alpha,C_\beta,\dots,C_\omega) $ is 
%
\[
\sum_{a=1}^K\sum_{b=1}^{K'} \dots \sum_{z=1}^{K'''} \\
 P(C_\alpha, C_\beta, \dots, C_\omega, a,b,\dots, z)\log P(C_\alpha, C_\beta, \dots, C_\omega, a,b,\dots, z).
\]

For a large number of clusterings or large K this quickly becomes inconvenient. In these cases we order the clusterings by adding new clusterings to the set based on maximizing the minimum pairwise distance to every other clustering currently in the set. This process is seeded with the informationally maximal pair within the set. This does not avoid triple-wise information overlap but works well in practice.  

\subsection{Physics Articles from arXiv.org}

ArXiv.org releases convenient metadata (title, authors, etc...) for all articles in their database. Additionally, a special set of 30 000 high energy physics articles are released with abstracts and citation networks. We apply our process to this graph of papers with edge types \textit{Titles, Authors Abstracts and Citations}. 

\begin{figure}[ht]
\centering
  \subfigure[]
  {
    \includegraphics[width=.43\textwidth]{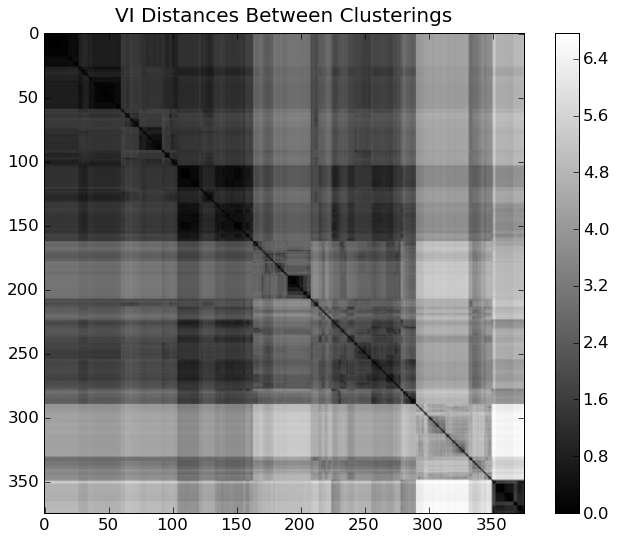} 
    \label{fig:arXivClusterSimMatrix}
  }
  \hspace{.1\textwidth}
  \subfigure[]
  {
    \includegraphics[width=.39\textwidth]{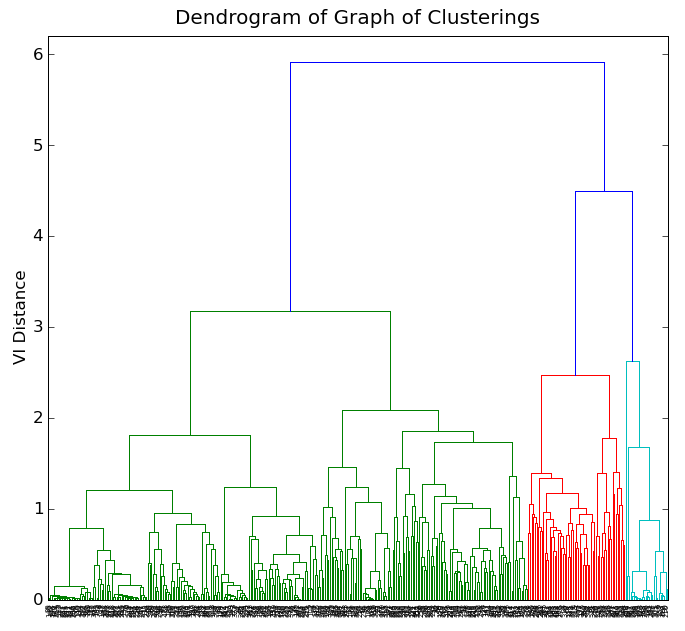}
    \label{fig:arXivClusterDendrogram}
  }
\caption{\small (a) The pairwise distances between the sampled clusterings form a graph. Note the dark blocks along the diagonal. These are indicative of tightly knit clusters. (b) A dendrogram of this graph. We use the ordering of the vertices picked out by the dendrogram to optimally highlight the blocks in the left image. } 
\vspace{-2ex}
\end{figure}

Articles are connected by \textit{title} or \textit{abstract} based on the cosine similarity of the text (using the bag of words model\cite{Blei2003}). Two articles are connected by \textit{author} by the number of authors that the two articles have in common. Two articles are connected by \textit{citation} if either article cites the other (undirected). We inspect this system with the following process discussed in greater detail above. 

These graphs are normalized by the $L_2$ norm and then the space of composite edge types is sampled uniformly.
That is 
$\omega_j = \sum_{i=1}^4 \alpha_i w_i$, where 
$ \alpha_i \in (-1,1) \ ,\   w_i \in \{\textrm{\textit{titles, abstract, authors, citation}}\} $.
The resulting graphs are then clustered using Clauset et al's FastModularity\cite{Clauset2004} algorithm. The resulting clusterings are compared in a graph which is then clustered to produce clusters of clusterings. The clusters of clusterings are averaged \cite{Strehl2003} and we inspect the resultant representative-clusterings.

The similarity matrix of the graph of clusterings is shown in Fig. ~\ref{fig:arXivClusterSimMatrix}. The presence of blocks on the diagonal imply clusters of clusterings. From this process we obtain  representative-clusterings. The various partitionings of the original set of papers vary considerably (large VI distance) yet exhibit high modularity scores implying a variety of high-quality clusterings within the dataset. 


\begin{table}[htb!]
\caption{\small Commonly appearing words (stemmed) in two distinct representative-clusterings. Clusters within each clustering correspond to well known subfields in High-Energy Physics (Traditional Field Theory/Lattice QCD, Cosmology/GR, Supersymmetry/String Theory). This data however does not show a strong distinction between the clusterings. Furher investigation is warranted.}
  \label{table:arXivWords}
  \begin{center}
  \begin{tabular}{ |c | c | }
  \hline
  Cluster & Statistically Significant Words in Clustering 1
  \\ \hline \hline
  1 & quantum, algebra, integr, equat, model, chern-simon, lattic, particl, affin \\ \hline
  2 & potenti, casimir, self-dual, dilaton, induc, cosmolog,  brane, anomali, scalar \\ \hline
  3 & black, hole, brane, supergrav, cosmolog, ads/cft, sitter, world, entropi \\ \hline
  4 & cosmolog, black, hole, dilaton, graviti, entropi, dirac, 2d, univers \\ \hline
  5 & d-brane, tachyon, string, matrix, theori, noncommut, dualiti, supersymmetr, n=2 \\
  \hline \hline
  \hline Cluster & Statistically Significant Words in Clustering 2 
  \\ \hline \hline
  1 & potenti, casimir, self-dual, dilaton, induc, energi, scalar, cosmolog, gravit \\ \hline
  2 & integr, model, toda, equat, function, fermion, casimir, affin, dirac\\ \hline
  3 & tachyon, d-brane, string, orbifold, n=2, n=1, dualiti, type, supersymmetr \\ \hline
  4 & black, hole, noncommut, supergrav, brane, sitter, entropi, cosmolog, graviti\\
  \hline
  \end{tabular}
  \end{center}
\end{table}

Analysis of this dataset is challenging and still in progress. We can look at articles in a clustering and inspect attributes like the country (by submitting e-mail's country code), or words which occur more often than statistically expected given the corpus. Most clusterings found show a separation into various topics identifyable by domain experts (example in Table \ref{table:arXivWords}) however a distinction between clusterings has not yet been
found. While the VI distance between metaclusterings presented in Fig.~\ref{fig:arXivClusterSimMatrix} is large it has so far proven difficult to identify the qualitative distinction for the quantitative difference. More in depth inspection by a domain expert may be necessary.

\section{Finding unexpected clusters}
\label{sec:nations}
 In many  applications, a domain expert can  predict  what the clusters will look like before computing a clustering. For instance,  if we have a graph whose vertices are countries with edges representing strong ties between the countries, we expect the geographic structure to have a strong influence on  the clusters for many  quantifications of a ``strong tie." While this domain specific information can be helpful in many ways, such a bias may obscure critical information about the data. For instance, we would like to know whether our graph has any good clusterings that are significantly different than a given list of clusterings, which  brings us to the topic of this section: 
{ \it Given a graph with multiple edge types,  and a set of clusterings,   how do we compute  an aggregate weighting function, such that  clusters of the aggregate graph   will be maximally different that the given set of clusterings?  } 
 
 This problem can be most naturally modeled as a multi-criteria optimization problem, where the first criterion will be the quality of the clustering and the second criterion will be  the  information independence  from  the given clusterings.   We need high quality clusterings, since we want the clustering to show  significant value.  One can impose a clustering structure  even to an  Erdos-Renyi style graph, but such clusters will not have any value.  We want  clusters that are statistically significant.  We also want to learn something that we do not already know, hence the second criterion, information independence. If there are multiple clusters  we want to deviate from, then  the second criterion will hold multiple criteria within. 
  
The methods we have proposed so far can be  employed to solve this problem as well.  Again, we need to quantify the information independence between two clusterings,   for which the variation of information metric, described in Section~\ref{sec:background:VI}, can be used.  And we need a method  to search the space for the  aggregate function efficiently, which can be done using the optimization methods  adopted in Section~\ref{sec:opt}.   

\vspace*{-2ex}  
\subsection{Case study: countries}
\vspace*{-1ex}
\label{sec:Nations}
We applied our methods to a  graph of the world where each node is a  country and modeled the relations between these countries with 6 types of edges:  
1) common land border,
2) major trade partners,
3) common language,
4) common religion,
5) common ethnicity,
6) similar labor force distribution (industry, agriculture, services, etc.).
%
\begin{figure}[t]
\centering
\vspace*{-1ex}
  \subfigure
  {
    \includegraphics[width=.75\textwidth]{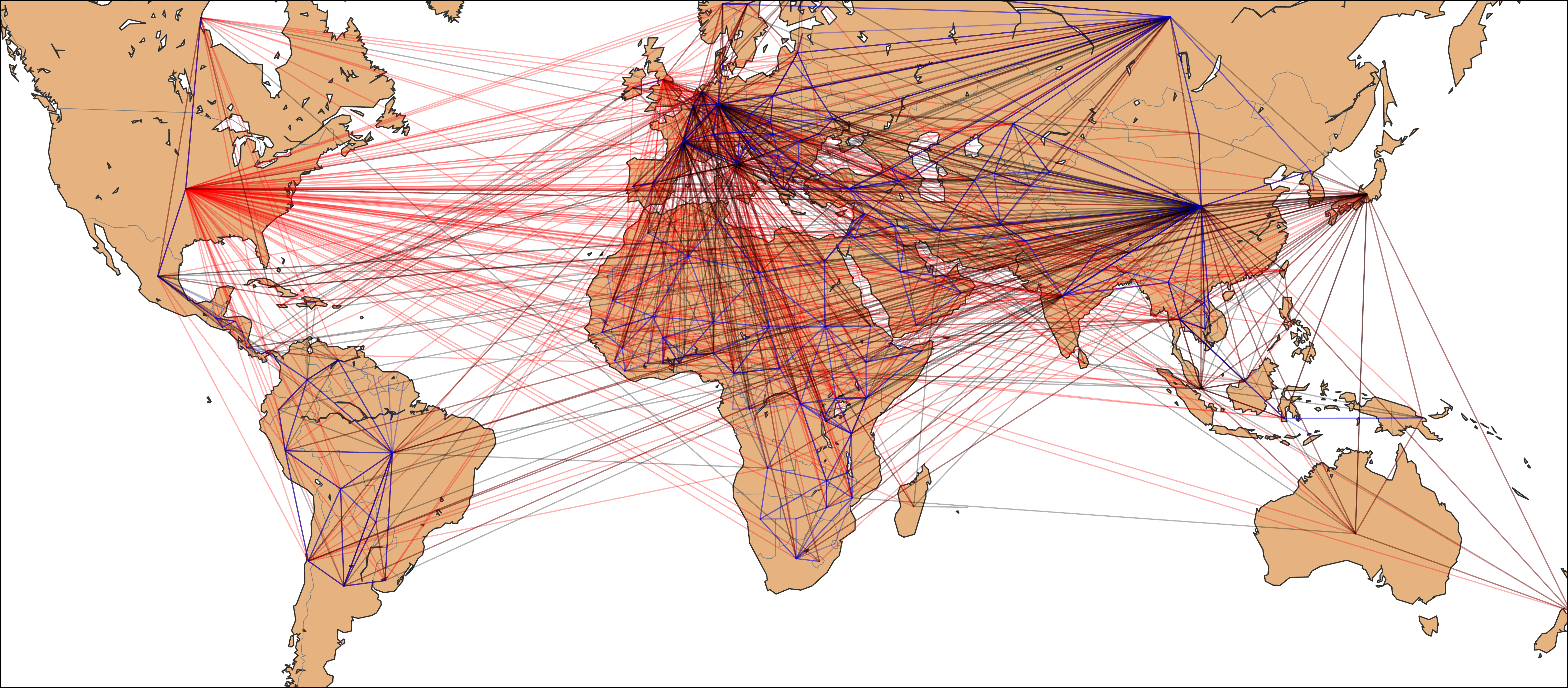} 
  }\\[-1ex]
  \subfigure
  {
    \includegraphics[width=.75\textwidth]{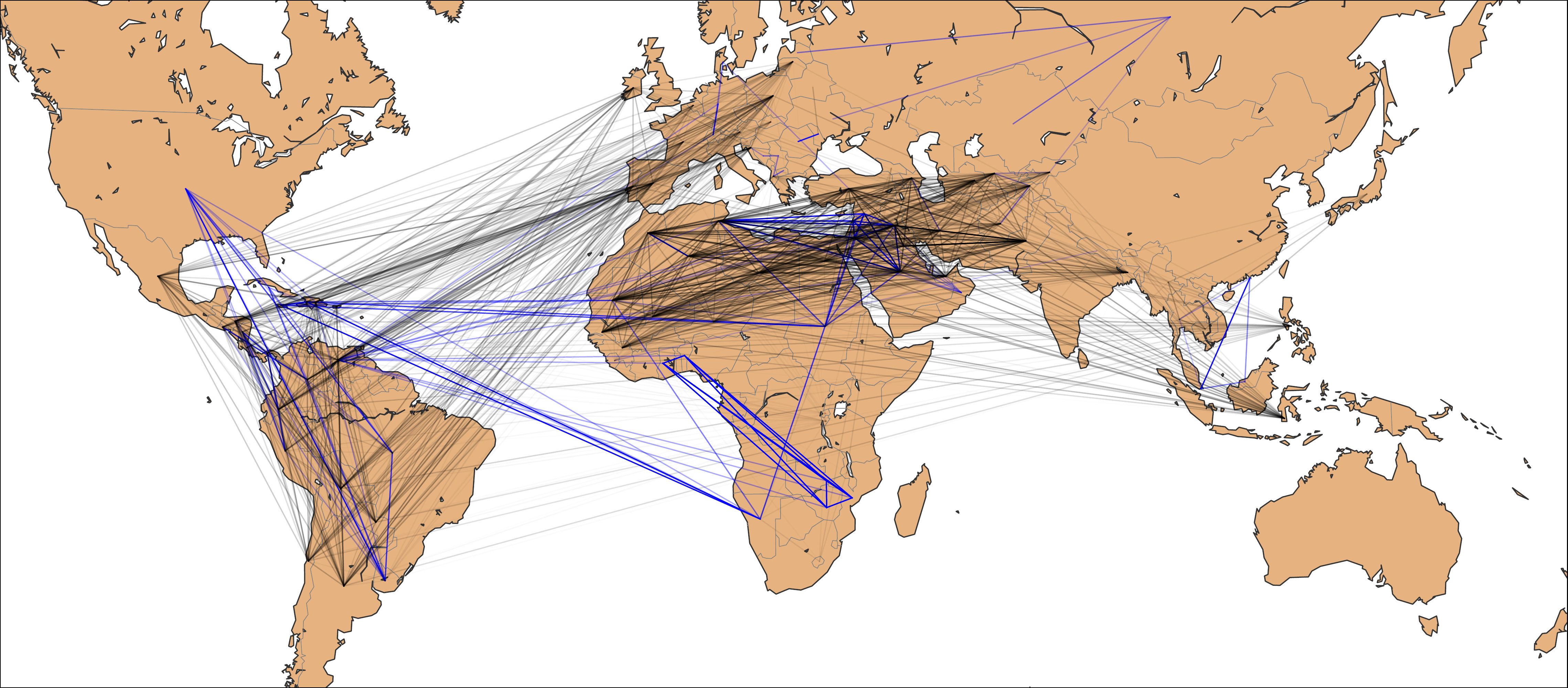}
  }
\caption{\small The world map with some edge types plotted. Top: Export/Import graphs (black and red) and physical land connections (blue). Bottom: Religion(black) and ethnicity (blue).} 
\label{fig:networkFigures}
\end{figure}
This data was taken from the CIA World Factbook 2006, which is conveniently available in structured text format~\cite{cia-nations}. 
%
The edge types borders and languages are unweighted while the others are weighted based on the probability that two people from these countries share the same trait. That is the probability that a Mexican citizen and American citizen share a common religion is .24, which becomes the weight of the edge.
The edge types of this graph are various in nature. Some are small world, some exhibit hubs and authorities, some have skewed degree-distributions, others do not (see Figure \ref{fig:networkFigures}).
Many of the edge types are correlated geographically. Ethnic, linguistic, religious, and trade relationships are often local. Can we find clusterings which are substantially different from this rule?  

\vspace*{-1ex}
\subsubsection{Searching with a linear aggregate function}
\vspace*{-1ex}
\label{sec:Nations:Linear}
As in Section \ref{sec:groundtruth} we optimized over the space of linear combinations of the edge types, but this time we allowed the weight of each edge $\alpha_i\in [-1,1]$, but we did not allow the weight of any edge to be negative, by truncating at  0.    
We used the HOPSPACK derivative-free optimization package and  a coefficient to balance between the two criteria. Clustering quality is given by the modularity metric and Clauset et al's software \cite{Clauset2004} is used to find the clusterings.

\begin{table}[thb]
\caption{\label{table:linearClusters}\small Groups of a clustering with maximum  information independence from continent-based clustering.  The  modularity of the clustering is 0.397, and the VI distance is 2.6.   The solution used  linear weights of labor(-0.99), exports (-0.5), language(0.5), imports (-0.61), religion(-0.06), borders (-0.95), and ethnicity(0.95)}
\begin{center}
\begin{tabular}{|c | c| }
\hline\hline
Cluster 1 &Angola, Belgium, Cambodia, Republic of the Congo, China, \\
&  Egypt, Greece, Hong Kong, Italy, Laos, Mali, Malaysia, Nigeria \\
& Rwanda, Singapore, Thailand, Taiwan, Papua New Guinea \\ \hline
Cluster 2 & Denmark, Mongolia, Macedonia, Turkey, Turkmenistan \\ \hline
Cluster 3 & Belarus, Georgia, Kyrgyzstan, Kazakhstan, Latvia, Lithuania\\
& Serbia, Russia,  Ukraine, Uzbekistan\\ \hline
Cluster 4 & Bangladesh, Benin, Canada, Cameroon,  Ireland, France, Ghana\\
& Germany, Haiti, Indonesia, India, Israel, Cote d'Ivoire, Jamaica \\ \hline\hline
\end{tabular}
\end{center}
\end{table}

In Table~\ref{table:linearClusters},  we present one of the clusterings we found.   We note that the modularity score  shows that  the clusters are statistically significant, and the distance from  the continents-based clustering is high, thus  this clustering provides new information.  
This solution strongly favors aspects like language, which is less geographically linked than others and strongly disfavors aspects like the sharing of a land border. The clusters obtained are at best vaguely related to the continental clustering.  One can see traces of  pre World War I  empires. 
Some of the clusters do highlight interesting linguistic groups that are conveniently spread across continental borders. 
Cluster 2 contains several countries with Turkic roots. Cluster 3 contains a Slavic family.  However, it is noteworthy that some countries with Turkic roots, such as  Kyrgyzstan, Kazakhstan, and  Uzbekistan, were grouped with the Slavic family. This shows that this high quality clustering  cannot be explained by language and/or ethnicity data alone. It is the combination that leads to this particular clustering.

\subsubsection{Drawback of linear aggregation }
\label{sec:Nations:Analysis}
In addition  to some interesting results, our experiments also pointed at some potential drawbacks of using a linear aggregate function.  We have observed that it is  possible to create clusterings  with significant information difference, but  these clusterings are not necessarily high in quality and/or statistically significant. And the reason for this is that linear aggregation  may yield dense or sparse  but Erdos-Renyi like random graphs. In this section, we will explain why this is the case. 
   
\begin{wrapfigure}{r}{.35\textwidth}
\vspace{-2pt}
\centering
\includegraphics[width=.35\textwidth]{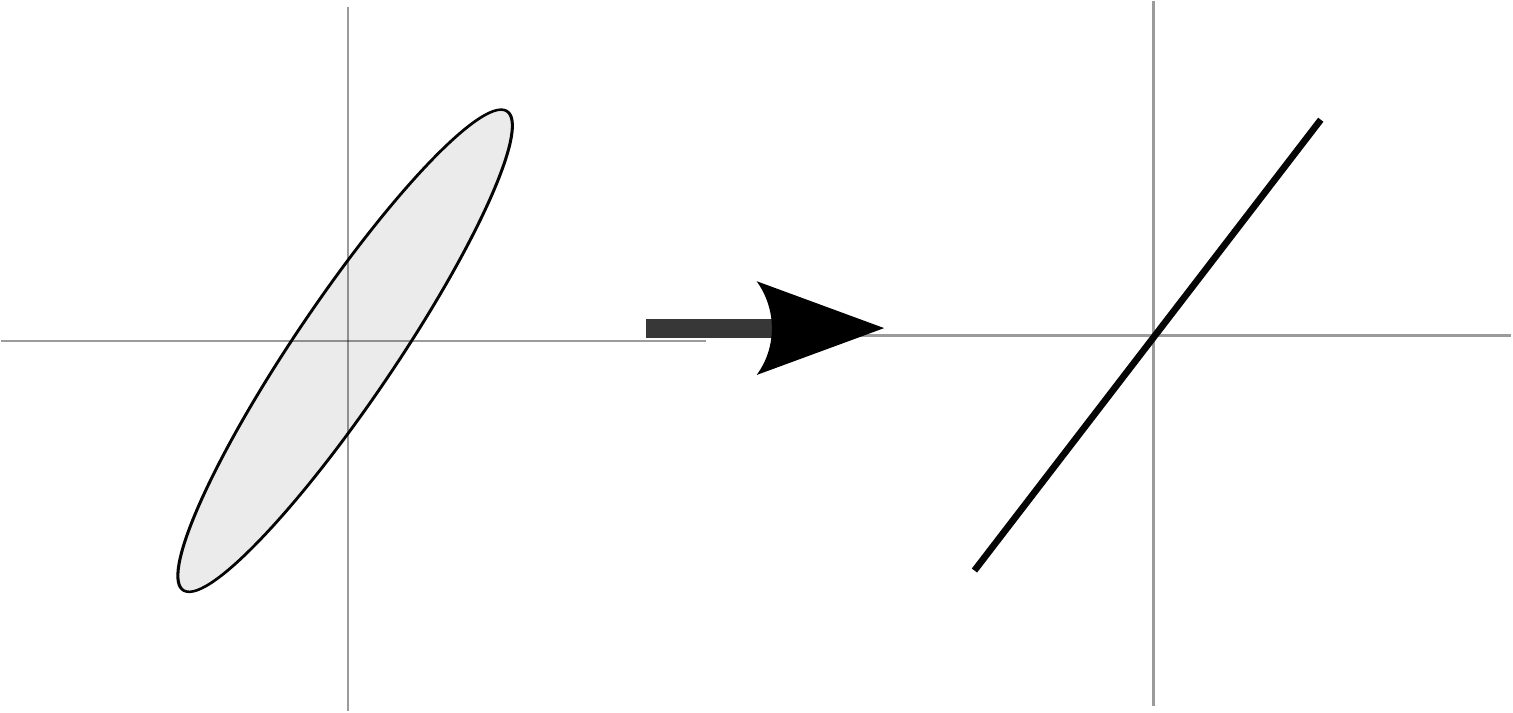}
\vspace{-1pt}
\caption{\small A clustering of a graph is just a low-rank approximation.}
\label{fig:ellipsoids_clustering}
\vspace{-1pt}
\end{wrapfigure}
The adjacency matrix of an undirected non-negatively weighted graph is symmetric  positive semi-definite,  and thus can be  pictured as ellipsoids with the major axes along the eigenvectors with the length of any axis  corresponding to the eigenvalue. If the matrix has few significant eigenvalues then the ellipsoid is flat in many directions and the matrix/graph can be simplified to a lower dimensional representation without much loss (see Figure~\ref{fig:ellipsoids_clustering}), which is the underlying idea of spectral clustering. A random or complete graph on the other hand is closer to  a symmetric sphere and cannot be reduced without substantial loss.

\begin{figure}[ht]
\centering
  \subfigure[The addition of two adjacency matrices produces a more uniformly symmetric graph]
  {
    \includegraphics[width=.45\textwidth]{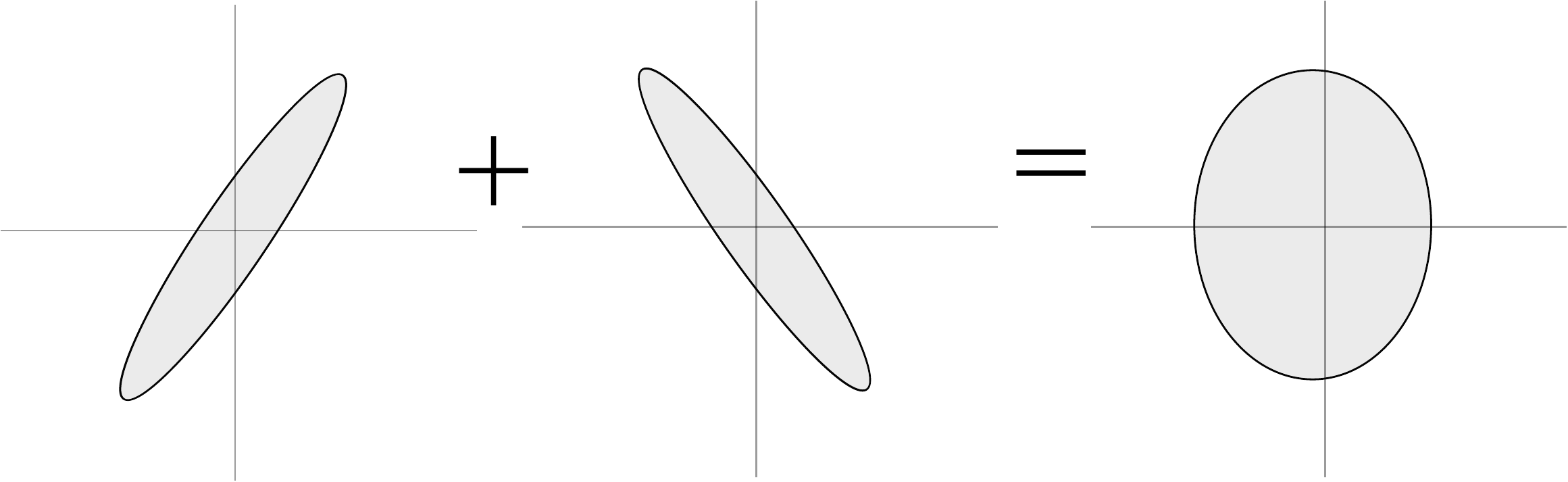} 
  }
  \hspace{.05\textwidth}
  \subfigure[The multiplication of two adjacency matrices produces a more degenerate, modular graph, amenable to low-rank approximations/clustering.]
  {
    \includegraphics[width=.45\textwidth]{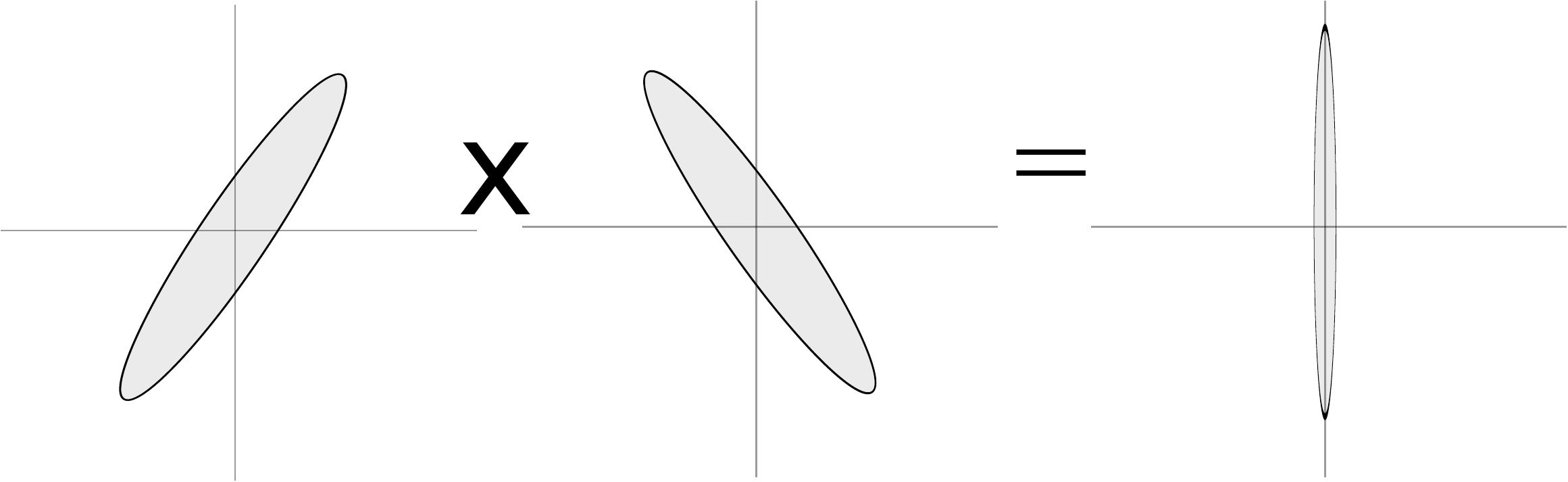}
  }
\caption{\label{fig:ellipsoids}\small  Illustration of effects of additive (a) and  multiplicative (b) aggregate function on the space of clusterings.   }
\end{figure}

This raises another interesting question: Can we use multiple edge types to improve the statistical significance of clusters?  For instance, a  multiplicative, as opposed to additive aggregate function may  restrict the search space (as illustrated in Figure~\ref{fig:ellipsoids}). In this case, we only take connections endorsed by more than one edge type into account. We demonstrate this with an example.  Clustering the countries graph with respect to only the religion and only language edge types produce modularity scores of 0.44 and 0.3 respectively. If we use a  graph where we have an edge if either a language or a religion edge exists,  the modularity score decreases to 0.23,  which means the clusters are statistically less significant. On the other hand, if we use a graph, where we have an edge only if we have both language and a religion edge, the modularity increases to 0.47, which means that the clusters have much more  significance. 
  
\subsection{Finding clusters with higher significance by using pair-products}
\label{sec:pair}
In the previous section, we discussed how  significance of clusterings  can be improved  by looking at products of edge types (i.e. we an edge only if both edge types support it). 
In this section, we present results based on this idea.  We  computed the modularity scores and  VI distances from the continental clustering for all single edge type and the product of each pair of edge types. We display ones with interesting results in Table \ref{table:pairs}.

\begin{table}
\caption{\small A few of the product-pairs and singleton edge types listed along with their modularity score and variation of information distance from the continental clustering. }
\label{table:pairs}
\begin{center}
\begin{tabular}{c | c | c || c | c | c}
\multicolumn{3}{ c||} {Single Metric} & \multicolumn{3}{c }{Metric Pair} \\\hline
Name & Modularity & VI distance &Name & Modularity & VI distance\\ \hline\hline
language  & 0.25 & 1.79 &  language $\times$ ethnicity & 0.43 & 2.40\\ 
religion  & 0.37 & 2.23 &  religion $\times$ ethnicity & 0.45 & 2.16\\ 
border & 0.42 & 0.69 & imports $\times$ religion & 0.41 & 2.17\\ 
exports & 0.29 & 1.94 & imports $\times$ labor & 0.29 & 1.85\\  
&& &  exports $\times$ imports & 0.37 & 2.35\\ 
&&& ethnicity $\times$ exports & 0.41 & 1.93\\  
&&& labor $\times$ language & 0.26 & 2.04\\ 
&&& labor $\times$ exports & 0.28 & 1.95\\ \hline \hline
\end{tabular}
\end{center}
\end{table}

Using the same  approach as in~Section \ref{sec:orderingInformationContent}, we select pairs that maximize set-wise information away from the continental clustering.  We found that the clustering that  is most different from the continental clustering is given by the product of the language and ethnicity graphs. Given those two clusterings the next most informationally distant comes from looking at import-export relationships. Finally, after the countries have been classified along these edge types the religion graph is the most distinguishing.  Note that  the products   yield  very high quality clusterings which are at the same time  distant from each other.

The three clusterings resulting from these three graphs are sufficiently distinct to distinguish almost all of the different countries from each other. The clusterings are sufficiently distant so that almost no two countries lie in the same cluster in all three of the clusterings. Equivalently most countries can be identified by their membership in the three clusterings. 
This achieves the goal set forth in Section~\ref{sec:latent} to efficiently represent a graph through well chosen representative clusterings, drastically reducing information with minimal content loss.

\begin{table}
\label{table:orderedGraphs}
\begin{center}
\begin{tabular}{| c | c | c | c | c }
Continents & Language $\times$ Ethnicity & Imports $\times$ Exports & Religion& ... \\ 
\end{tabular}
\caption{An set of edge types ordered to yield maximum set-wise information / minimum set-wise redundancy. Each in turn is as different as possible from each of the preceding clusterings. Note that these represent relatively orthogonal ideas (geography, culture, economy, ...) in turn.}
\end{center}
\end{table}

Another interesting result in our experiments was the groups of countries that were  always grouped together, which are listed in Table~\ref{table:productClusters}. This list will not be surprising to many, but  it still underlines the concept of metric invariant clusters in graphs with multiple edge types.  

\begin{table}
\caption{\label{table:productClusters} \small Some groups of countries that were consistently clustered together} 
\begin{center}
\begin{tabular}{l | l}
\hline
Group 1 & Lebanon, Syria, Yemen \\ \hline
Group 2  & Norway, Sweden \\ \hline
Group 3 & Belgium, Denmark, Netherlands \\ \hline
Group 4 & Colombia, Dominican Republic, Ecuador, El Salvador, Guatemala \\
& Honduras, Mexico, Nicaragua, Peru, Venezuela\\ \hline
\end{tabular}
\end{center}
\end{table}

\vspace{-2ex}
\section{Conclusions}
\label{sec:conc}
\vspace{-2ex}
We addressed clustering in the context of graphs  with multiple edge types.  We investigated  several problems within this context: recovering an aggregation scheme from ground-truth clustering,
 finding a meta-clustering structure and efficiently representing this structure, and finding unexpected  clusters. We also presented case studies on real data sets. 
 
 The main result of our work is that  working on graphs with multiple edge types prevents loss of crucial information. Instead of a single clustering, a rich clustering structure can exist with clusters of clusterings, and latent clusters can be discovered that compactly explain the underlying graph.  
 We showed that  high quality clusters which are significantly different than expected can be found, and significance of clusters can be improved by looking at intersections of multiple edge types.  
Another important conclusion of  this work is that,  despite the increased complexity due to multiple edge types, we have the algorithmic tools to make the associated  problems tractable. 

We hope that  our work will draw the attention of the research community to this important problem, which bears many  intriguing research challenges, and we can see much room for growth in this topic. We have mostly  focused on linear  aggregation and nonlinear combinations   (as we have seen in Section~\ref{sec:pair})  should be investigated.  More intelligent sampling methods should be possible. And most importantly, working on more data sets  will give us a chance to better evaluate our methods.  
  
\small
\singlespacing
\bibliographystyle{plain}
\bibliography{matt,ali}

\end{document}